\documentclass[10pt,conference]{IEEEtran}
\IEEEoverridecommandlockouts

\usepackage[switch]{lineno}
\usepackage{cite}
\usepackage{float}

\usepackage{amsmath,amssymb,amsfonts}
\usepackage{algorithmic}
\usepackage{graphicx}
\usepackage{url}
\graphicspath{{./figs/}}
\usepackage{textcomp}
\usepackage{xcolor}
\usepackage{caption}
\usepackage{subcaption}
\usepackage{xspace}
\usepackage{enumitem}
\def\BibTeX{{\rm B\kern-.05em{\sc i\kern-.025em b}\kern-.08em
T\kern-.1667em\lower.7ex\hbox{E}\kern-.125emX}}

\newcommand\tool{\textit{Kub}\xspace}

\begin{document}

\title{Kub: Enabling Elastic HPC Workloads\\on Containerized Environments\thanks{This research is supported by the European Commission
under the Horizon project OpenCUBE (GA-101092984).}}

\author{\IEEEauthorblockN{Daniel Medeiros, Jacob Wahlgren, Gabin Schieffer, Ivy Peng}
\IEEEauthorblockA{\textit{Department of Computer Science}\\
\textit{KTH Royal Institute of Technology, Sweden}\\
\{dadm, jacobwah, gabins, bopeng\}@kth.se}
}

\maketitle

\begin{abstract}
The conventional model of resource allocation in HPC systems is static. Thus, a job cannot leverage newly available resources in the system or release underutilized resources during the execution. In this paper, we present \tool, a methodology that enables elastic execution of HPC workloads on Kubernetes so that the resources allocated to a job can be dynamically scaled during the execution. One main optimization of our method is to maximize the reuse of the originally allocated resources so that the disruption to the running job can be minimized. The scaling procedure is coordinated among nodes through remote procedure calls on Kubernetes for deploying workloads in the cloud. We evaluate our approach using one synthetic benchmark and two production-level MPI-based HPC applications -- GROMACS and CM1. Our results demonstrate that the benefits of adapting the allocated resources depend on the workload characteristics. In the tested cases, a properly chosen scaling point for increasing resources during execution achieved up to $2\times$ speedup. Also, the overhead of checkpointing and data reshuffling significantly influences the selection of optimal scaling points and requires application-specific knowledge.
\end{abstract}

\begin{IEEEkeywords}
HPC, Cloud, scaling, Kubernetes, Elasticity, Malleability
\end{IEEEkeywords}

\section{Introduction}
In recent years, two main trends have contributed to the rising importance of the convergence between High-performance computing (HPC) and cloud. The first trend is the increased compute resources on recent HPC systems, where a single node may be equipped with multiple high-end CPUs, GPUs, and 200-500 GB RAM. For instance, the Frontier supercomputer has four GPUs and 1TB memory per node. Therefore, the conventional coarse-grained static resource allocation strategy on HPC systems faces challenges in resource utilization, and efforts in exploring fine-grained dynamic resource allocation strategies that have matured on the cloud are increasing. Secondly, workloads are evolving towards more complex patterns, e.g., workflows~\cite{ejarque2022enabling,medeiros2023gpu} incorporating traditional scientific simulations, and machine learning and in-situ data analytics. Meanwhile, the availability and accessibility of compute resources in public cloud providers like Amazon, Google, and Microsoft have attracted users to explore running HPC applications on the cloud, and previous limitations of scaling up HPC applications on cloud infrastructure are being addressed in recent active research development~\cite{beltre2019enabling, misale2021towards, milroy2022one,araujo2023libcos}.

As user-facing cloud services tend to experience spikes or cyclical demand in their time series, cloud-native workloads, such as search and web serving, have shifted their development from a monolithic to a microservice, loosely-coupled structure. This shift was made to leverage the native adaptive autoscaling capabilities that are built into many deployment systems. These scaling decisions may be grounded in models that take into account application-specific metrics, such as the incoming number of requests, response time, available system resources (i.e., CPU, memory), and historical patterns~\cite{ROSSI2020161}. 

Elasticity, which refers to the degree to which a system can adapt its resource provision to workload changes \cite{herbst2013elasticity}, has become a promising direction for converged cloud and HPC computing in recent years. In HPC systems, resource allocation to jobs is coarse-grained and static, so a job with high peak usage of resources may have to wait a long time until all resources become available for its execution. From a resource management perspective, this can block spare resources in an HPC system until a large task can start. In contrast, cloud-based systems offer fine-grained dynamic resource allocation to cloud-native workloads. This is because the cloud approach for resource allocation has developed and matured over the years to support the demand for elastic execution from cloud-native workloads. Specifically, HPC users and developers are seeking opportunities in elastic execution for high computing power, high availability, and cost efficiency.

Certain types of HPC workloads, such as deep learning, can make extensive use of this elasticity. This owes to the fact that such workloads and related applications (i.e., TensorFlow and PyTorch) are already structured in a way that shares the training/predicting load in batches among the active nodes. This makes solutions such as Elastic Horovod\footnote{Elastic Horovod: https://horovod.readthedocs.io/}, which detects new nodes in real-time, feasible to use. However, the same cannot be directly applied to tightly-coupled workloads that are built atop Message Parsing Interface (MPI) for distributed-memory programming. Traditional HPC schedulers, such as SLURM, do not support the dynamic change of nodes in their resource allocation to a job~\cite{huber2022towards}. Once a job starts running, adding new node resources is cumbersome, if not impossible. This means that a full start/stop process is necessary for changing the number of nodes, and the previously allocated resource will be lost, and time should be spent waiting for a new allocation if that was not previously available. 

This work aims to address the problem of elastic scaling of HPC workloads by reusing the already provisioned infrastructure, with a focus on the cloud-containerized environment. We achieve this through the usage of the de-facto industry standard container orchestrator Kubernetes and multiple representative HPC workloads that allow the understanding of the gap from achieving elastic execution.

In this paper, these are our major contributions:
\begin{itemize}
    \item We identify technical challenges of running MPI-based applications with a state-of-art container orchestrator.
    \item We propose an approach for elastic horizontal scaling of tightly-coupled workloads, in particular ones that are using MPI in a containerized context. 
    \item We implement the proposed approach and quantify the overhead and its benefits in the HPC applications GROMACS and CM1.
    \item We discuss the application requirements and the trade-offs of elastic scaling of HPC applications in containerized environments. 
\end{itemize}

This paper is structured as follows. Section II discusses the current state-of-art means for executing HPC workloads on the cloud, including some cloud-first technologies that are later used in this paper. Section III shows our approach for elastic scaling workloads and the details of our implementation. Section IV dives into the specifics of the applications workloads we are using for this work, while Section V displays our setup and results. Section VI briefly discusses some related works while Section VII consists of the conclusions and our future works.
\section{Background}
In this section, we describe the differences between HPC and Cloud workloads and introduce the building blocks for enabling elastic execution in this work.

\subsection{HPC and Cloud Workloads}

Cloud computing applications tend to be loosely-coupled and fault tolerant \cite{5599083}. User-facing applications should be able to scale up and down according to demand, and techniques such as load balancing, where the processing batch can be directed to a less-stressed node, helps with the design of such applications. Some of the representative cloud workloads are search, data streaming, web serving, and in-memory databases \cite{mulia2013cloud}. 

Meanwhile, high-performance computing workloads tend to be tightly-coupled as there is usually an interdependence between the calculations being distributed among the nodes. For instance, the CORAL-2 benchmark suite contains representative HPC workloads\footnote{Coral-2 benchmarks: \url{https://asc.llnl.gov/coral-2-benchmarks}}, which includes molecular dynamics, quantum Monte Carlo, fluid simulation and cosmology.

The Message Parsing Interface (MPI) is the dominant approach for communication in HPC workloads, as a means for performing distributed calculations over a large number of compute nodes. In many cases, these calculations are explicitly or implicitly blocking operations because their execution time is determined by the slowest node due to data dependencies. Although MPI has introduced some mechanisms recently to support dynamically adding or removing members to a communicator, the schedulers on HPC systems have little support to change node allocation to a job once it starts. Therefore, in practice, this means that the only way of changing the number of allocated nodes during the execution is through the process of restarting the application.

\subsection{Containers and Orchestrators}
Containers are a solution to isolate the resources of tasks executing in the same node. In Linux-based systems, the isolation of resources is implemented through the \texttt{cgroups} feature. Docker is the industry standard for containers; however, due to security concerns (i.e., the container runs as root by default), other container technologies such as Singularity \cite{10.1371/journal.pone.0177459} and Podman \cite{10.1007/978-3-030-59851-8_23} have been more used in HPC systems. In the case of Docker, the container is defined by a Dockerfile which contains the base image (i.e., an operating system, such as Alpine Linux or Debian) to be used and a series of deterministic commands related to the application that one desires to deploy - this includes the installation of dependencies, compilation of software/libraries and setting environment variables. 

These images are deployed by orchestrators, which are responsible for not only distributing the containers among the nodes, but also monitoring them and ensuring that their characteristics - such as minimum number of replicas - stay consistent with the desired number by the user. Some of the popular orchestrators are Kubernetes, OpenShift Container Platform and Docker Swarm, with all of them being tested in HPC environments \cite{marathe2019docker}. 

A Kubernetes cluster consists of at least two nodes, one being the master and the other being a worker node. The former runs decoupled applications responsible for the communication interface between the user and the cluster (named api-server), the scheduler, an etcd object storage for storing data and metadata from the cluster and a controller manager. Inside every worker node, there is an application named kubelet, responsible for dealing with incoming requests from the api-server, such as executing pods and exposing pod metrics for scrappers.

The basic unit within a Kubernetes cluster is the pod - an abstraction of resources for executing a container. Computational resources are defined as the time used by the CPU and the Memory. A worker node may contain one or more pods at the same time, and similarly, a pod may have one or more containers executing within the same resource domain. A set of equal pods can be associated as a ReplicaSet, where the controller manager ensures that the desired number of pods will be running in case of failure. Deployments are an extended ReplicaSet with more useful features. 

Architecturally, the idea of a pod was developed for tasks that should run for an indefinite amount of time. For time-limited, finite tasks, the concept of jobs is used. In practice, a job is marked as completed when a certain number of pods have been executed and successfully finished. In this case, the pods do not necessarily need to execute and finish at the same time, although in some cases (e.g., MPI-based workloads) they do.

By default, each pod has its own IP inside the Kubernetes network and can freely communicate with other pods unless otherwise defined. Aside from the IP, a pod can reach others through the usage of the built-in domain name server records which is deterministic and based on variables such as the name of the pod and its namespace. However, as either the IP or the DNS record might be mutable, Kubernetes introduces the concept of service as a means to gather a set of pods providing similar service (i.e., user-facing) and, at the same time, exposing it outside the network.

Any kind of Kubernetes-defined or custom-defined resources are defined through the usage of YAML files. The schema used by such files through Kubernetes is standardized and checked for errors before execution. Upon acceptance of the YAML file, the default Kubernetes scheduler first looks for feasible nodes and then, among the returned group of nodes, checks for the most viable ones through scoring. 

Finally, the security mechanism in Kubernetes is performed through a system called RBAC (role-based access control). One may define a certain role that is able to get, create, delete or edit certain types of resources and then assign such role to, for example, a pod. By default, a pod is not able to create other pods or modify cluster-level settings.

\subsection{Autoscaling}
In cloud settings, autoscaling means changing the amount of resources of an existing allocation. This may be through the change of already allocated CPU and memory (in this case, vertical scaling) or through the increment or decrement of the number of available nodes for the application (horizontal scaling).

Kubernetes provides a built-in vertical pod autoscaler (VPA) and a horizontal pod autoscaler (HPA) by default. However, as these tools were originally designed for cloud applications, there are some problems when trying to use them with HPC workloads. First, the VPA consists of three components, namely the admission controller, recommender and updater; based on the historical pattern of CPU/Memory usage of the application, the recommender outputs a value for CPU and Memory. If the recommended value is too different from the used one, the pod is evicted and restarted. This is because Kubernetes currently does not allow changes in the requested resources of a pod unless the pod is restarted. While there are some ongoing works to address this limitation and change the allocated CPU/Memory resources dynamically, this is currently not in production versions. 

The built-in HPA queries the resource utilization periodically and according to user-defined policies, such as the threshold for a certain metric, it decides to scale up or down. The metrics can be either directly related to physical resource usage (i.e., CPU or Memory) or application-level metrics, such as the number of incoming requests. However, in tightly-coupled workloads, the built-in horizontal scaler has no effect at all. If one is running an MPI application, for example, the newly started rank would execute the same calculations by itself, being unable to join dynamically the already existent communicator. 

\subsection{Volcano and MPI} 
Volcano\footnote{Volcano: \url{https://volcano.sh}} is a batch system for Kubernetes, providing tools for certain types of workloads that run on frameworks such as TensorFlow, Spark and MPI. Kubeflow\footnote{Kubeflow: \url{https://www.kubeflow.org/}} is another framework, with a strong focus on machine learning, that enables one to run MPI workloads on the cloud.

The YAML file for deploying a Volcano MPI job consists in defining a \texttt{VolcanoJob} with two types of pods: one named \texttt{master} and another that has a \texttt{worker} suffix. Additionally, the YAML file also specifies two plugins that are necessary to run MPI workloads, the \texttt{svc} and the \texttt{ssh} plugins. The former is responsible to enable all the pods within a job to visit each other by domain name and \textit{by default} establish a policy of not allowing any other pod outside service to communicate with that network of pods. Additionally, it creates a list of all the working pods that will execute. The second plugin generates a key pair locally and mounts \texttt{/root/.ssh} as a read-only directory at every Volcano-created pod. This ensures that all the pods will have \texttt{ssh} passwordless authentication between each other, a prerequisite to smoothly run MPI jobs. There is currently an ``MPI plugin" for Volcano which can be used instead of the \texttt{svc} and \texttt{ssh} plugins, in practice, replacing both, but not allowing the same degree of flexibility required by this paper. 

OpenRTE~\cite{castain05:_open_rte}, part of the Open MPI project, is the heart of how Volcano executes its MPI jobs. As its name implies, OpenRTE is a runtime environment that provides services related to process management, communication coordination and resource allocation. When launching a process through the \texttt{mpiexec} wrapper, one may specify which hosts are necessary to connect - this is collected by Volcano, and OpenRTE interacts with a daemon at every node - named \texttt{orted} - through a defined communication protocol to coordinate the launch and information such as launch path, environment variables and command line arguments. OpenRTE and orted keep an interaction for message passing and monitoring until the end of the execution. 

\begin{figure}[t]
\includegraphics[scale=0.065]{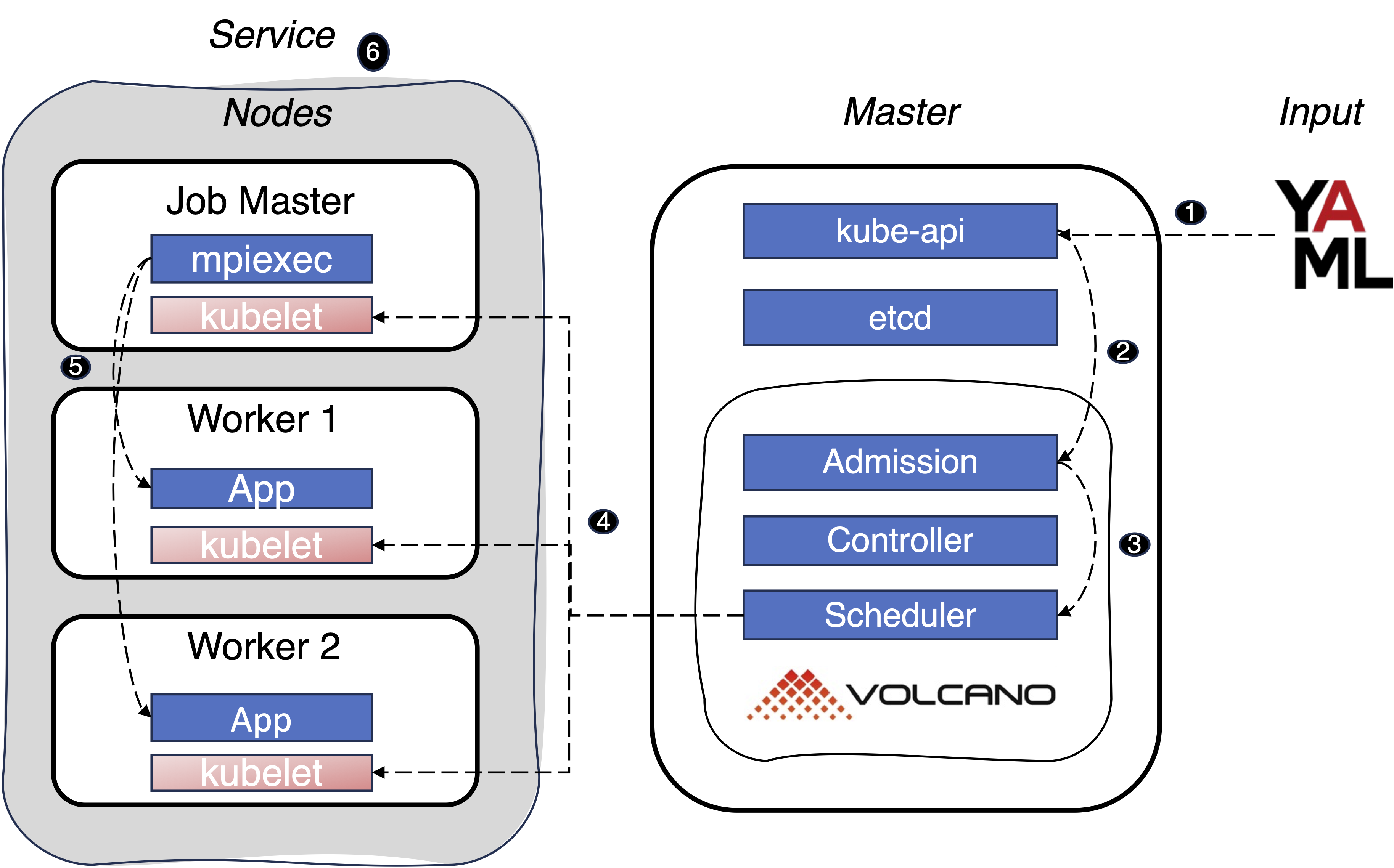}
\centering
\caption{The overall structure of a Volcano deployment in Kubernetes. I)  Defined through a YAML file, the job is communicated to the api-server (or kube-api). II) This call is intercepted by Volcano's admission controller that checks whether the YAML has all required fields and contains no error. III) If everything is correct, data is sent to the scheduler which verifies whether it is possible to allocate the job. IV) When the resources are available, the kubelets inside each node are ordered to create the pod allocation with the desired containers for the job. A pod can contain one or more containers, and nodes can also run more than one pod. V) The ``master'' pod awaits all ``worker'' pods to be active and open their \texttt{ssh} daemon. When they are ready, orted starts the MPI job among them. VI) All the pods are encapsulated by a ``service'' type, so they can communicate using each other by domain. Volcano's scheduler and controller, responsible for monitoring the jobs, effectively replace the ones included in Kubernetes by default.} \label{fig:kube-struct}
\end{figure}

Figure \ref{fig:kube-struct} displays the structure of a Volcano system. The master pod is responsible to start the MPI jobs, while the workers execute the workload itself. In general, the \texttt{worker-0} pod is considered the root rank of the MPI execution. While the spawning process of containers might take longer for some,  the master node will keep crashing and restarting until \texttt{orted} can successfully connect with all the listed nodes. After the execution, the job is marked as Complete at the Kubernetes cluster.


\subsection{Resource Monitoring} \label{sec:kub-monitoring}
The kubelet that runs on every node collects information regarding CPU and Memory through cAdvisor\cite{tolaram2022cadvisor}, a profiling application maintained by Google that is intended for Docker containers. The Prometheus Operator\footnote{Prometheus Operator: \url{https://prometheus-operator.dev/}} is a Kubernetes application that is able to scrape metrics not only from all the kubelets, but also from the master node, exposing them for further usage in JSON format. By connecting to the web API of Prometheus, one is capable of retrieving the time-series and history of the metrics, as well as checking the current status of a node in near-real time. Other applications that are combined with Prometheus are Grafana (for visualization) and the Elastic stack (Kibana and Elasticsearch). Jaeger\footnote{Jaeger: \url{https://www.jaegertracing.io/}} is another library for instrumenting and exposing application-level metrics in cloud applications, that can also be scraped by Prometheus.



\section{Methodology}
As to enable tightly-coupled workloads to be elastic, we introduce \tool, an extension to Kubernetes written in Python.

\subsection{Architectural Design} \label{sec:arch}
\tool is composed of three major components: the Monitor, the Coordinator and the Executors. We choose to use these terms as a means to separate from Kubernetes terminology (i.e., master and workers). The Monitor is responsible for deciding when to scale as well as the creation of new pods. We understand that the decision to scale should be application-specific and left to the developers as there are trade-offs regarding the restart of a job, as it will be discussed in Section \ref{sec:eval}. Due to the need for the privileges for pod creation, the Monitor may be deployed either inside the Kubernetes realm or outside. When deployed as a Kubernetes pod, it is necessary to modify the RBAC permissions from the Monitor pod so it is able to use the Kubernetes API to deploy other pods. If outside the Kubernetes realm, it is necessary to ensure that it has enough access to do so, usually granting it access to the configurations present in the host system.

In a nutshell, \tool works by being the process \#1 when the pod is started. The rationale behind that is because, in traditional MPI applications running on Kubernetes, where Volcano is used, the launch of a \texttt{mpiexec} application as process \#1 means that the job is deeply related to the status of the job, failing or completing according to it. Here, we use \tool for coordinating the resources for an application restart, to avoid time wasted in restarting all the infrastructure. 

The checkpointing and restarting procedure is application-specific and should be written according to the application's needs. First, during checkpointing, we observe two major patterns. Some production-level HPC applications designed for long runs have already had support for checkpointing the files upon the receiving of a SIGTERM signal. Others checkpoint at each user-defined time interval. For the restarting procedure, it is necessary to handle the change of parameters in input files or a change of parameters in the command line to specify which checkpoint file should be used.

The Coordinator runs at the master pod in Volcano and effectively acts as a gRPC server. Its main purpose is for coordinating the launching of MPI applications and the eventual restarting procedure (i.e., it does not perform calculations), thus it is very lightweight. The Monitor acts as a gRPC client and its role is to define when the criteria of when the scaling process will take place, to create new pods and to tell the Coordinator how many new pods should be expected.

In a similar fashion, the executors are the worker pods as described by Volcano, or it can also be a newly-created pod by the Monitor (which we define here as a "scaling pod"), also being gRPC clients to the Coordinator. Every timestep (usually 10 seconds), they check the status of the job with the coordinator. This is done to avoid the automatic completion of the job when the application is paused, which can be due to checkpointing in some cases.

\begin{figure*}[t]
\includegraphics[scale=0.45]{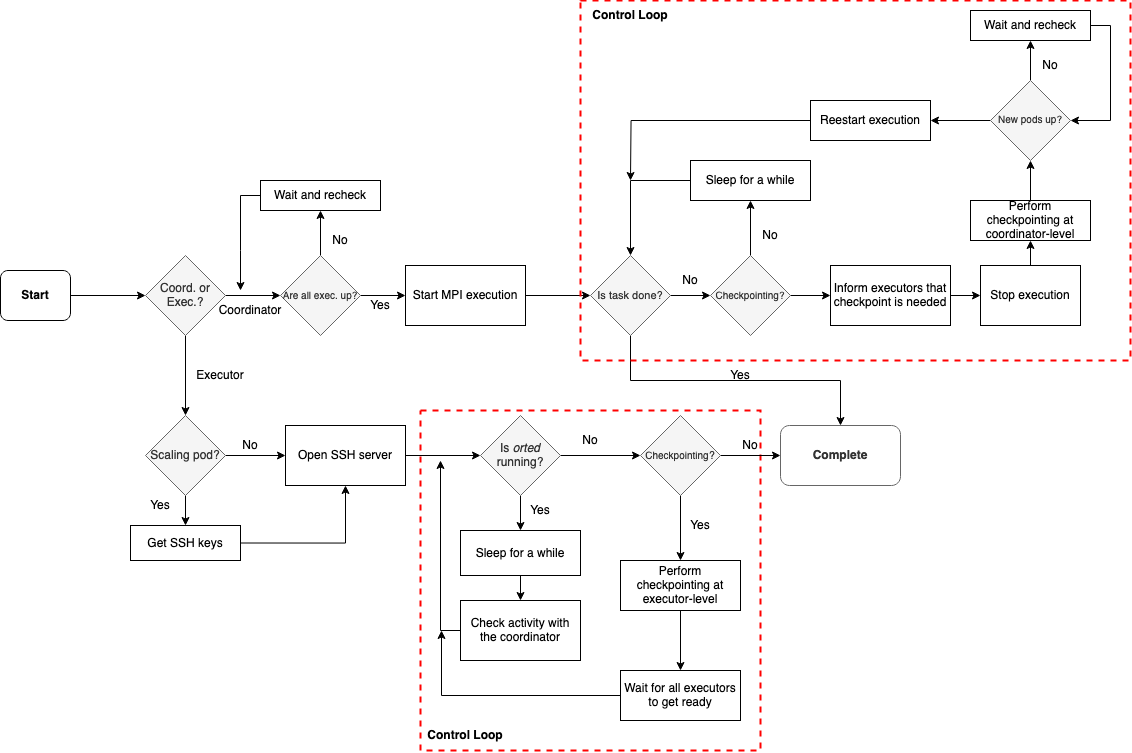}
\centering
\caption{The decision flowchart of \tool when executing. One of the major benefits of Kub is to be able to use non-provisioned infrastructure as any newly-created pod can decide to join the others by exchanging SSH keys with the Coordinator. Furthermore, the control loops ensure that the scaling can be performed multiple times during the application's execution time. } \label{fig:bia-flowchart}
\end{figure*}

\subsection{Elasticity}
During the horizontal scaling of the application, when one or more pods are intended to be added, it is necessary to coordinate among all the already existing ones. The Monitor sends a message to the coordinator with the intended number of nodes to scale. As executors check the job status with the coordinators periodically, the latter changes (figuratively) its own job status to ``Scaling''. With such a message, the executors keep waiting for the checkpointing and the scaling process to be done instead of finishing the execution. The monitor then proceeds to create the necessary number of pods.

Figure \ref{fig:bia-flowchart} displays the entire flowchart when performing horizontal scaling. The newly-started pods cannot enter into the MPI network by default as orted cannot reach them without their IP addresses. Thus, when an additional pod is starting, it sends a message to the coordinator with its IP address and requests the current key pair shared among all the pods and, upon received, the scaling pod uses it to allow the passwordless login through orted. The coordinator keeps track of the received IP and adds it to the list of available pods to run. Finally, the job can be restarted. The process repeats when it is necessary to increase the number of pods again.

\subsection{RPC Calls}
There are many information exchanges between the master node and the monitor or the worker nodes. In this work, we use gRPC\footnote{gRPC: \url{https://grpc.io/}} -- a library developed and maintained by Google which aids in the process of sending and dealing with RPC calls between applications. In practice, gRPC works through the concepts of protocol buffers: a protocol that enables serialization/deserialization of messages in many languages. By writing the message code directly into the protocol language, the message can be converted (and thus used) by languages such as Python, C++, Javascript and Go. All gRPC clients include a stub, which contains all the available remote procedures and is used to send a command to the server through a channel (usually a single HTTP 2.0 connection), and concurrent calls may be multiplexed into that channel. The server has threads waiting for any connection and will handle all necessary commands according to the procedures written in the code, returning a message thereafter. Table \ref{table:rpccalls} displays all the calls, along the parameters, that are used in \tool.

\begin{table*}[t]
\centering
\caption{List of RPCs and its parameters that are used in \tool}
\begin{tabular}{|c|c|c|c|}
\hline
\textbf{Call Name} & \textbf{Parameter(s)} & \textbf{Direction} & \textbf{Description} \\ \hline
Scale & \begin{tabular}[c]{@{}c@{}}Number of Nodes,\\ Mode of Scaling\end{tabular} & Monitor $\Rightarrow$  Coordinator & \begin{tabular}[c]{@{}c@{}}This call tells the Coordinator that there are available\\  resources and that the Job should get ready to scale.\end{tabular} \\ \hline
RetrieveKeys & Name of Node & Scaling pod(s)  $\Rightarrow$  Coordinator & \begin{tabular}[c]{@{}c@{}}A newly-started pod can ask the Coordinator for its public and\\ private key as to establish a ssh connection for orted.\end{tabular} \\ \hline
JobInit & Name of the Node & All pods  $\Rightarrow$  Coordinator & \begin{tabular}[c]{@{}c@{}}This is used by the executing pods to tell the coordinator pod \\ that the current pod is alive and ready for execution.\end{tabular} \\ \hline
activeServer & (None) & All executors  $\Rightarrow$  Coordinator & \begin{tabular}[c]{@{}c@{}}This checks whether the master is alive and whether \\ the working node should do any client-side checkpointing.\end{tabular} \\ \hline
checkpointing & (None) & All executors  $\Rightarrow$  Coordinator & \begin{tabular}[c]{@{}c@{}}This is used to confirm that the checkpointing \\ was done by the pods.\end{tabular} \\ \hline
endExec & (None) & All executors  $\Rightarrow$  Coordinator & \begin{tabular}[c]{@{}c@{}}This is used to confirm that the execution is about to \\ finish on the active executors.\end{tabular} \\ \hline
\end{tabular}
\label{table:rpccalls}
\end{table*}

\subsection{Deployment}
The deployment of \tool is done through the usage of a launcher script, written in Python, that is used for both the Coordinator and the Executors. Based on the standardized hostname, the launcher discerns the difference between each other and branches the code. The Monitor is also packaged as a Python script and can be run either outside or inside a pod, as described by Subsection \ref{sec:arch}.

We focus on the horizontal scaling in this work as a first step for enabling elastic MPI-based HPC applications. The vertical solution was envisioned for single-node workloads that rely on, for example, OpenMP for thread parallelism. Our future work will investigate the vertical scaling. The horizontal scaling enabled by \tool is described as follows. When running an HPC workload, a pod should already default to use the maximum available resources available from a node. If it is not available, one can use the horizontal scale to increase the number of MPI ranks, with the new one using the spare resources at the moment.  

\subsection{Coordinating pods}
Aside from the elastic scaling, there is an extra benefit to our approach. In Volcano, the Master pod has the \#1 process running \texttt{mpiexec}, while the workers execute a \texttt{ssh} daemon. A problem happens when the Master pod executes before all workers are up. When this happens, the master pod crashes due to not being able to reach all specified pods. The master pod should then be restarted, in which it will re-execute its command in the hope that all pods will be active.

\tool avoids this problem by having the Coordinator acknowledge all the initial executors as active before starting the \texttt{mpiexec} application. As the Coordinator knows the number of initial executors, it will also expect to receive a similar amount of \texttt{JobInit} RPC calls. In the case of the Executors trying to communicate before the gRPC server is up, the fault tolerance is handled in such a way that it will retry sending the message until a response from the Coordinator is received.
\section{Applications} \label{sec:apps}
In this section, we characterize all the target applications for this work. In particular, there is a focus on their checkpointing process and how they were adapted to be used together with \tool.

\subsection{CM1}
The Cloud Model 1 (CM1) \cite{ABenchmarkSimulationforMoistNonhydrostaticNumericalModels} is a numerical model for idealized studies of the atmosphere with a focus on deep precipitating convection. It is actively maintained by the National Center for Atmospheric Research (NCAR).

Spanning over 230,000 lines in total, CM1 is written in Fortran and supports either OpenMP, for shared memory computations, or MPI, for distributed memory. The input for the application is done through the usage of a \texttt{namelist.input} file which contains a very large number of parameters regarding the model to be simulated.

In this paper, we use the default workload for CM1 that is provided with the default \texttt{namelist.input} file. Similarly, we chose to use MPI for the calculations. For MPI, the root rank is the one responsible for distributing the load among all the available ranks and dealing with I/O operations, which includes checkpointing and writing the final results as well. In practice, these characteristics allow CM1 to start the calculation with certain numbers of ranks and finish it with a different number.

\textbf{Checkpointing.} Every predefined number of timesteps, the root rank of CM1 outputs multiple checkpoint files that start with the prefix \texttt{cm1rst\_} followed by the number of the checkpoint in cardinal order. The usage of this checkpointed file should be explicitly handled by the \texttt{namelist.input} file through the \texttt{irst} parameter.

\textbf{Algorithm.} The application-specific algorithm in \tool consists in checking whether the root rank is executing the code. If it is, then it iterates over the entire checkpointing directory and its related files (that starts with
\texttt{cm1rst\_}), looking for the most recent checkpoint based on the filename. With the id found, a small text operation for replacing the \texttt{irst} parameter on the \texttt{namelist.input} is done. The application is then ready to restart with a new number of ranks.

\textbf{Modifications.} There were no modifications in the vanilla code of CM1 aside from the Makefile being changed for the selection of the OpenMPI compiler wrappers.

\subsection{GROMACS}
GROMACS \cite{abraham2015gromacs} is an open-source software suite for molecular dynamics simulation. One of its major popular use is as the backend for the distributed protein folding in the Folding@Home\footnote{Folding@Home: \url{https://foldingathome.org/}} project.

As a command line application, GROMACS is built entirely in C++ and supports a wide range of parallel and accelerating technologies, such as OpenMP and its built-in threading, MPI and GPUs (through the SYCL library). There is also support for SIMD intrinsics such as AVX-256 and AVX-512. 

At the core of GROMACS is the \texttt{mdrun} engine, responsible for not only executing molecular dynamics calculations but also stochastic dynamics and energy minimization. It takes a wide range of parameters as input. It is important to mention that, due to the intrinsic randomness of the calculations, two GROMACS simulations are unlikely to yield the same results (although both of the results will be correct), even after stopping and resuming the same simulation. 

GROMACS has been previously tested in a cloud setting \cite{doi:10.1021/acs.jcim.2c00044}, in particular through the usage of the AWS heterogeneous clusters (ARM, Intel, AMD CPUs and different GPUs) scattered over the world, managed with Hyperbatch and aided by the Elastic Fabric Adapter for communications between nodes located in different regions. S3 was used for storing intermediate files.

As a workload for this paper, we use one of the molecular dynamics benchmarks provided by the Max Planck Institute for Multidisciplinary Sciences\footnote{benchMEM: \url{https://www.mpinat.mpg.de/grubmueller/bench}}, namely the \texttt{benchMEM} (82 000 atoms, protein in membrane surrounded by water) benchmark.

\textbf{Checkpointing.} The checkpointing in GROMACS is handled automatically when it receives a SIGTERM signal, writing the files as soon as it is received and gracefully stopping the application. The root rank is responsible for writing the checkpointing files and also initially reading them, distributing the data among all the available ranks. 

\textbf{Algorithm.} The application-specific algorithm consists in sending a SIGTERM to GROMACS when it is time for checkpointing, waiting for it to write the files and killing the application. To re-execute the application from checkpointed data, an additional flag is introduced into the running command.

\textbf{Modifications.} No modifications in the GROMACS code were performed for this paper. The application was built according to its documentation, with the flags to build using MPI and to use its own FFTW. 
 
\subsection{PARINT benchmark}
We design PARINT, a \underline{p}arallel distributed benchmark with configurable \underline{ar}ithmetic \underline{int}ensity. Arithmetic intensity is the number of arithmetic operations per byte loaded from memory and measures the balance between compute and memory demands in an application~\cite{ofenbeck2014applying}. PARINT consists of an outer loop over an array, with a variable number of operations per array element determined by the parameter \verb|NLOOP|. With \verb|NLOOP=1|, the arithmetic intensity is low and the workload is bound by the available memory bandwidth, while a high value of \verb|NLOOP| gives a high arithmetic intensity, scaling with available compute power. PARINT is implemented in C and parallelized using MPI.

\textbf{Checkpointing.} We implement checkpointing in PARINT upon receiving a SIGUSR1 signal. Upon the arrival of this signal, PARINT will checkpoint and gracefully exit. On startup, PARINT checks whether a checkpoint file exists and loads it into memory before continuing with the main loop. The cost of checkpointing depends on the data size, determined by the \verb|ARRAY_SIZE| parameter.

\textbf{Algorithm.} The algorithm consists in propagating a SIGUSR1 signal to PARINT upon receiving the call for increasing the resources. PARINT then checkpoints and the coordinator waits for the new pods to start before restarting the execution.
\section{Evaluation}\label{sec:eval}

\subsection{Infrastructure}
In this study, we use a single-node cloud testbed that consists of an Intel i7-7820X processor, with 8 cores (16 logical cores) in total, and 32 GB of DDR4 memory at 2133 MHz. In terms of storage, the system contains an Intel Optane SSD 900p with 480 GB, a Kingston UV400 SSD, 2x Seagate Barracuda with 2 TB each and a Samsung EVO NVMe driver with 1 TB, which the operating system (Ubuntu 22.04) is running. This cloud testbed also includes an Intel I219-V single-port 1 gigabit Ethernet controller.

Furthermore, we use Kubernetes v1.23 which is deployed through k3d\footnote{k3d: \url{https://k3d.io}} as it emulates a single-node system. k3d includes by default the services for domain name resolution (CoreDNS), networking (traefik) and monitoring (metrics-server). Furthermore, Volcano v1.7.0 was deployed in that system. 

By using a single node to run multiple pods, possible delays due to communication are mitigated and the focus shifts to the methodology itself.

All the applications were compiled using GNU Compiler Collection (GCC) v11.3 together with OpenMPI 4.1 when it was necessary as a dependency. Python 3.10 was used both on containers and system-wise for running our launcher.  

\begin{figure*}
\includegraphics[scale=0.6]{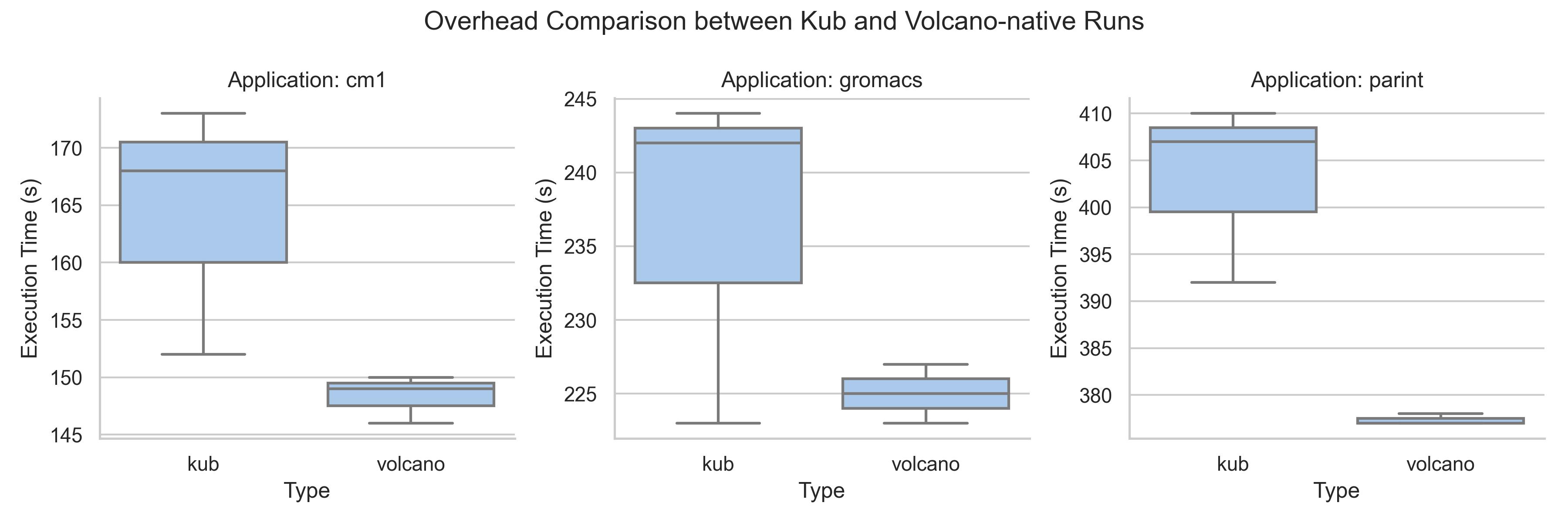}
\centering
\caption{Calculated overhead of the applications used in this work. Label \tool means that the applications were started using the custom launcher that coordinates the scaling, although no scaling was performed, while "Volcano" is the traditional way of using MPI applications on Kubernetes. For this experiment, each application was executed using three MPI ranks, one per Kubernetes pod.} \label{fig:overhead}
\end{figure*}

\subsection{Containerization of Applications} 
The process of building images for applications is widely documented, thus this paper will not discuss in-depth such process. We use Debian 11 ``slim" as a base image for our containers, and build them using in two stages. The slim version of Debian removes many files related to documentation and language support, allowing the image to weigh roughly 80 MB (in comparison to the 125MB from the full image).

The two stages process consists in having the first image to compile the application itself with all the necessary building tools and development libraries. With that done, the compiled executable is transferred to the second image, which will contain only the necessary runtime libraries for execution.  

\begin{figure}
    \centering
    \includegraphics[width=\linewidth]{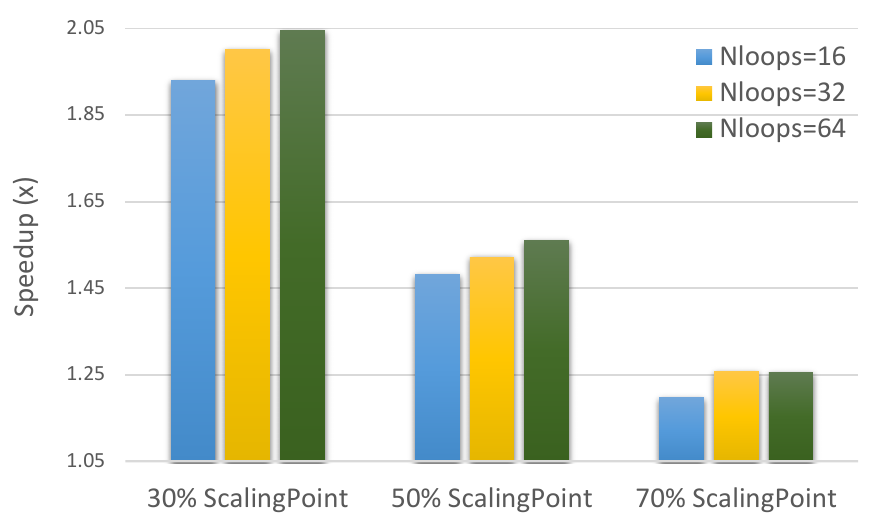}
    \caption{Sensitivity test of increased compute intensity and benefit from scaling up from 2 to 6 ranks at three scaling points $30\%$, $50\%$, $70\%$, respectively.}
    \label{fig:sensitivity}
\end{figure}

\subsection{Overhead of \tool} \label{sec:ovh}
We evaluate the impact of using our approach (i.e., a launcher) in comparison to the vanilla version of the applications in Kubernetes. For this study, we did not perform any type of scaling as we wanted to measure the effects of i) the coordination among the pods and ii) the effect on thread-sharing of running a gRPC server and client in the system. 

The timings were extracted from the total job duration from start ("Running" status) until completion ("Complete" status), and the boxplot in Figure \ref{fig:overhead} displays the results and also the conditions in which the application was executed.

The results show that there is a slight (between 10 to 15\%) overhead for running \tool instead of the vanilla application in Kubernetes. However, this overhead can be largely attributed to the fact that the launcher sleeps between timesteps, so an action that should be performed during the time that the launcher is sleeping will be performed only at the beginning of the next timestep - which includes checking whether all executors are alive and/or even ending the execution for checkpointing. 
\begin{figure*}
\centering
\begin{subfigure}[b]{0.325\linewidth}
\centering
\includegraphics[width=\textwidth]{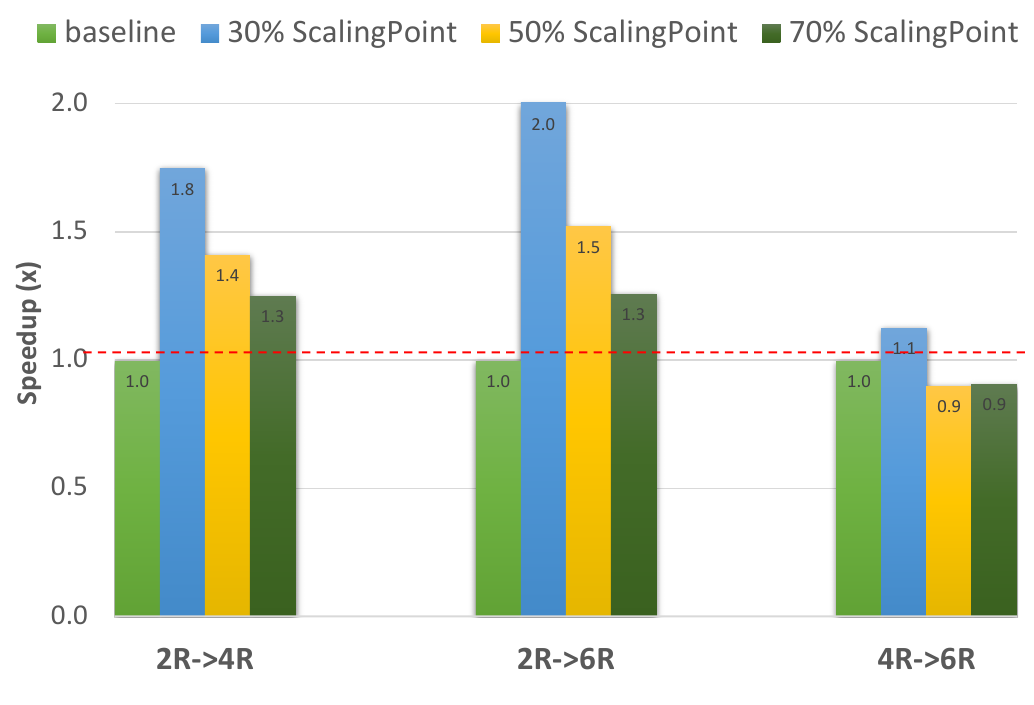}
\caption{PARINT}
\end{subfigure}
\begin{subfigure}[b]{0.325\linewidth}
\centering
\includegraphics[width=\textwidth]{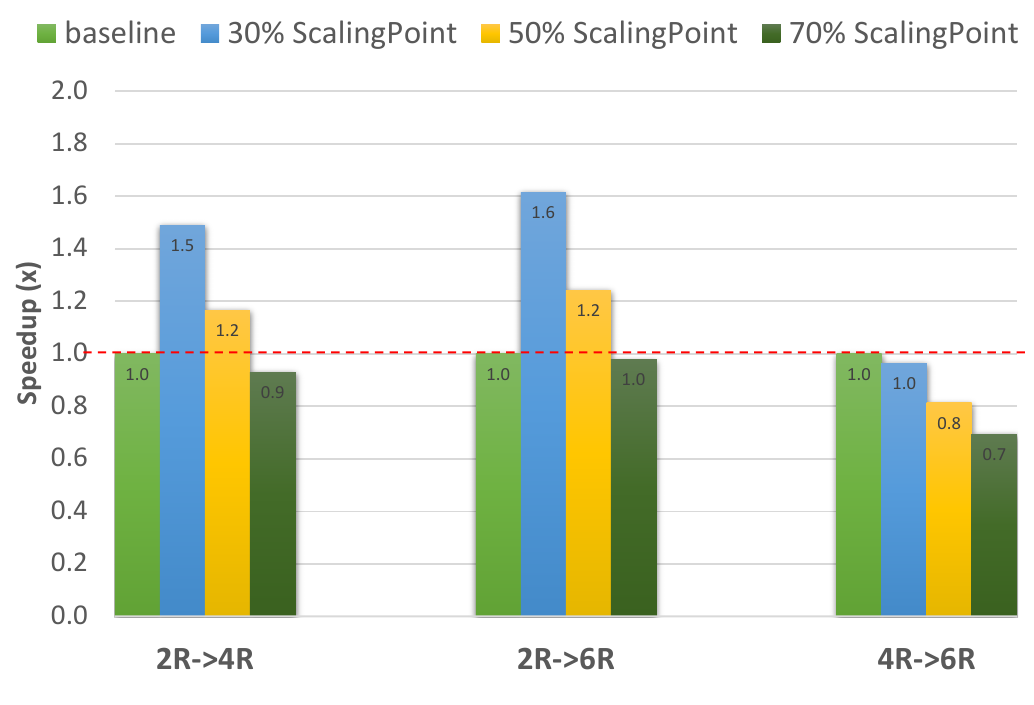}
\caption{CM1}
\end{subfigure}
\begin{subfigure}[b]{0.325\linewidth}
\centering
\includegraphics[width=\textwidth]{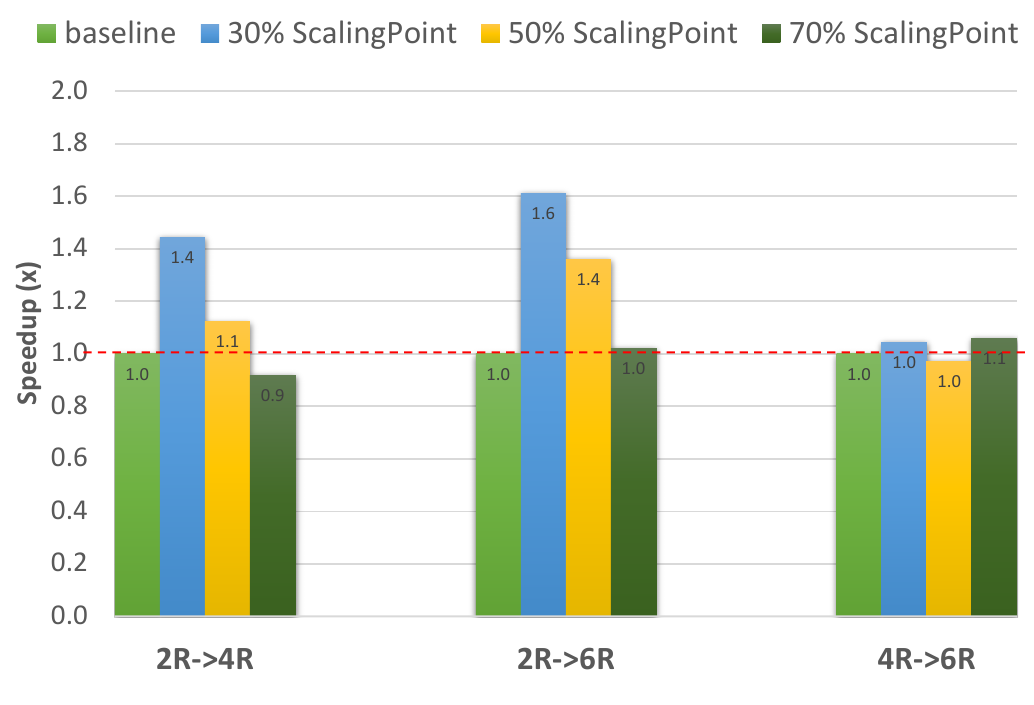}
\caption{GROMACS}
\end{subfigure}
\caption{The results for the elastic scaling performed by this work. Each case was executed 3 times, totalling 36 experiments per application. The label on the X axis refers to the \textit{amount} of resources that is being introduced into the application, while the colours for each bar refer to \textit{when} the scaling was performed. Refer to Sections \ref{sec:hscale} and \ref{sec:disc} for an extensive discussion about this figure.} \label{fig:results}
\end{figure*}
\subsection{Scaling Experiments Overview}
Sections \ref{sec:sens} and \ref{sec:hscale} display results that are used for different analyses. However, all the experiments were designed and performed following similar procedures. 

In particular, applications described in Section \ref{sec:apps} were used to study the effects of elastic scaling. As previously stated, the applications were not modified for these experiments; instead, an understanding of its checkpointing and resuming procedures was necessary and implemented into the launcher script. 

In such experiments, the only factor when deciding to scale was time; there was no monitoring of available resources due to the hardware constraints - rather we assume that such resources will be available at a certain point of the execution of this application. 

The speedup is calculated over a baseline time, which is a vanilla run of the application (i.e., without scaling). Each of the experiments discussed above had their transition between ranks from one state to another at 30, 50 and 70\% of the baseline time. This is done to investigate the influence of \textit{when} the scaling is done. For the scenarios where the starting point is 2 MPI Ranks, the baseline is a full execution of 2 MPI Ranks; an analogue situation happens when the starting point is 4 MPI Ranks.

\subsection{Sensitivity Test} \label{sec:sens}
This experiment uses PARINT to measure how applications with low or high computational intensity might benefit from the elastic scaling at different scaling points. We change the available \texttt{NLOOPS} parameter, where 16 makes the benchmark more memory-bound and 64 makes it more compute-bound. Figure \ref{fig:sensitivity} displays the obtained speedup over a baseline with no scaling. As the scaling point increases, the speedup decreases because there is less gain from scaling, but equal overhead from checkpointing. As for arithmetic intensity, the speedup increases as the intensity is increased showing that compute-bound tasks benefit more from parallelism than memory-bound tasks in our setup. Overall, a memory-bound task would have to scale out earlier in the execution to benefit, while a compute-bound task can scale out later in the execution.

\subsection{Horizontal Scaling} \label{sec:hscale}
In these experiments, we evaluate CM1, GROMACS and PARINT (with the NLOOPS parameter equaling 32) with our methodology, and the results can be seen in Figure \ref{fig:results}. 

For each application, we executed three different scenarios, all of them dealing with the increase of the available resources. The first scenario deals with an increase of 100\% of resources (2 to 4 MPI ranks), the second one is a 200\% increase (2 to 6 MPI ranks) and the last scenario is an increase of 50\% (4 to 6 MPI ranks). For each of these scenarios, we evaluated the speedup at different moments of increasing the amount of resources. The relationship among the factors is mostly reconfirmed, and it is possible to observe one more relationship between the amount of resources and the speedup.

\subsection{Discussions}\label{sec:disc}
There are two major insights to be drawn from the results shown in Sections \ref{sec:ovh}, \ref{sec:sens} and \ref{sec:hscale}.

\begin{enumerate}
    \item The decision of scaling or not depends on the amount of resources and how much time the application has expected to finish its execution.
    \item Although there is overhead from using \tool, the possibility of scaling might enable the mitigation of it.
\end{enumerate}

The first point is clearly illustrated by the results. In general, all three applications behave similarly: at the scenarios with the scaling point at 30\% and 50\% of the baseline time, there are improvements in the execution time for the case for scaling from 2 to 4 ranks and when scaling from 2 to 6 ranks, with the latter being usually faster than the former as the amount of resources is increased. Where there is scaling from 4 to 6 ranks, there is a perceived slowdown due to the increase of resources not being high enough to compensate for the time to stop for checkpointing and restarting. 

However, this is not the case at 70\% scaling point. Instead, there is a perceived slowdown on the application, meaning that if the application had run that much, it is better to let it finish instead of doing all the coordination for restarting.

That said, one question that arises is related to which applications are feasible to apply this methodology. As seen in Section \ref{sec:sens}, PARINT is an ideal case and the speedup gains increase according to the arithmetical intensity of the application. In real applications, such as GROMACS and CM1, such gains are limited by the amount of non-computing operations (i.e., I/O) that are performed during its execution. Furthermore, such applications might also be able to stop and restart with a different number of ranks, also distributing the remaining load among the existing nodes. 

Finally, in relation to the second point, the 10 to 15\% of overhead that is shown between Volcano and \tool can be mitigated if a proper speedup is obtained with the resource increase - in particular, because the speedups increase ranges from 30 to 80\%, as seen in Figure \ref{fig:results}.

\section{Related Works}
The emergence of converged cloud and HPC computing has attracted increased works in understanding the feasibility and gaps of scaling. We classified the literature related to this paper into four categories as follows.

\textbf{Feasibility and infrastructure.} Several works \cite{saha2018evaluation, beltre2019enabling,marathe2019docker,abraham2020use} discussed the impact of using containers for HPC workloads, and evaluate orchestrators such as Docker Swarm and Kubernetes for such cases. In some cases, the latency impact of using InfiniBand and TCP/IP protocols is measured as well. Malleability is also proposed in MPI and PMIx~\cite{huber2022towards}. Liu et al. \cite{liu2021performance} evaluated the impact of multi-tenancy in different types of containers (Docker and Singularity), considering both UMA and NUMA types of hardware, and reaching the conclusion that MPI applications suffer some degree of degradation due to each container being provided with its own networking namespace, with this effect being mitigated for applications that don't have much interprocess communication.   

\textbf{Malleability.} There are several ongoing works in malleability for HPC applications. In particular, MPI Sessions was extended to support dynamic resource allocations \cite{huber2022towards}. Some parallel programming languages support a change in the number of nodes. In Charm++, for example, an interface named Converse Client Server sends and receives signals related to the expansion or reduction, and these signals can either be internal (the application takes its own decisions) or from an external application.


\textbf{Scaling of HPC Workloads on the cloud.} There is a trend of extending the Kubernetes scheduler to support HPC workloads better. In particular, Misale et al.~\cite{misale2021towards} proposes a scheduler for Kubernetes called KubeFlux based on the ideas from Flux~\cite{ahn2020flux}. Using NFD, KubeFlux incorporates heterogeneous awareness for different compute resources. Milroy et al.~\cite{milroy2022one} further contributed an MPI Operator and the Fluence plugin to Kubernetes, demonstrating scaling HPC applications up to 3000 MPI ranks on IBM Cloud and AWS.


\textbf{Performance measurements and analysis.} Gupta et al.~\cite{gupta2014evaluating} evaluated the performance and cost of selected HPC applications across multiple HPC and Cloud platforms. They focus on identifying suitable HPC workloads running on the cloud and proposed optimizations to Cloud virtualization mechanisms to match the characteristics of HPC workloads. Sukhija et al.~\cite{sukhija2019towards} discussed the requirements of a monitoring tool in HPC environments and proposed the integration of a tool called OMNI (from NERSC) with current state-of-art tools that are used in cloud computing settings, such as Prometheus and Grafana.



\section{Conclusion}
In this paper, we proposed a methodology for elastic scaling of tightly-coupled HPC workloads on the cloud. Our evaluation shows that the obtained speedup heavily relies on the quantity of resources to be introduced on the system and also at which point the scaling will be done, as there is a tradeoff between checkpointing overhead and the benefits of additional resources.

We show that the underlying mechanism for coordination between MPI processes on containerized environments is complex and deals with different technologies and software to turn the idea into reality - gRPC for coordination, Kubernetes for resource management, SSH and MPI for running tasks through the network, Volcano as a monitoring aid for the tasks, plus the application-level knowledge for ensuring that the checkpointing works. Furthermore, our work can be advanced in two future fronts of work.

The first front is monitoring (Section \ref{sec:kub-monitoring}). The current work does not leverage resource awareness but rather builds a fixed time model for simplicity (i.e., at 30\%, 50\% and 70\% of a base execution time). For the elastic scaling to be effective in a production-level system, two factors should be considered: i) how many resources will be available on the system to an application, and ii) how much can the application benefit from additional resources. This is widely studied in cloud environments, especially when dealing with quality of service for users.

The second front is to analyse more complex patterns of scaling. This work only scaled up once although \tool can do it multiple times through the same algorithm. However, designing experiments and analyzing results for such patterns is difficult as there is a large space to explore. Finally, as one can scale up for performance, we think that scaling down can also play a big role in resource management and energy consumption in the future.

\section*{Acknowledgments}
This research is supported by the European Commission
under the Horizon project OpenCUBE (GA-101092984).


\bibliographystyle{IEEEtran}
\bibliography{main} 

\end{document}